\documentclass[aps,prl,twocolumn,showpacs]{revtex4}

\usepackage{amssymb}
\usepackage{graphicx}

\newcommand{\text}[1]{\mbox{#1}}
\newcommand{\revision}[1]{#1}

\begin{document}

\title{Self-Excited Multifractal Dynamics}

\author{V. Filimonov}
\email[]{vfilimonov@ethz.ch}
\affiliation{ETH Zurich, Department of Management, Technology and Economics, Zurich, Switzerland\\ and State University --- Higher School of Economics, Nizhny Novgorod, Russian Federation}

\author{D. Sornette}
\email[]{dsornette@ethz.ch}
\affiliation{ETH Zurich, Department of Management, Technology and Economics, Zurich, Switzerland}

\date{February 24, 2011}

\begin{abstract}
We introduce the self-excited multifractal (SEMF) model,
defined such that the amplitudes of the increments of the process are
expressed as exponentials of a long memory of past increments. 
The principal novel feature of the model lies in the 
self-excitation mechanism combined with exponential nonlinearity, 
i.e. the explicit dependence of future values of the process on past ones.
The self-excitation captures the microscopic origin of the emergent endogenous self-organization
properties, such as the energy cascade in
turbulent flows, the triggering of aftershocks by previous earthquakes
and the ``reflexive'' interactions of financial markets. The
SEMF process has all the standard stylized facts found in financial time series, which are robust
to the specification of the parameters and the shape of the memory
kernel: multifractality, heavy tails of the distribution of
increments with intermediate asymptotics, zero correlation of the
signed increments and long-range correlation of the squared
increments, the asymmetry (called ``leverage'' effect) of the
correlation between increments and absolute value of the
increments and statistical asymmetry under time reversal.
\end{abstract}



\maketitle


Strong scientific efforts are aimed at characterizing,
understanding, predicting and using the rich intermittent nonlinear
dynamics of extended natural and social systems. In the last two
decades, significant progress has been achieved through the
development of new concepts and tools, in particular those of
scaling, multi-scaling and multifractals that emphasize the role of
the symmetry of scale invariance, the coexistence of and interplay
between multiple scales and the hierarchical organization of fractal
singularities. In this vein, the multifractal formalism provides
powerful metrics to quantify the complex spatio-temporal
fluctuations occurring in such diverse systems as hydrodynamic
turbulence (velocity increments and energy dissipation)
 \cite{Frisch}, seismicity (stress field and earthquake triggering)
\cite{SornetteOuillon}, financial systems (asset returns)
\cite{Lux08}, biology (healthy human heart-beat rhythm) \cite{Ivanov} 
and hydrology (river runoffs) \cite{Kantelhardt2006}.

Multifractality reflects the presence of both long-range dependence
and hierarchical organization at many scales. A
particularly important realm of application is to model and 
explain the long-memory processes occurring
in the time domain \cite{Beran94}, such as found in
hydrodynamic turbulence \cite{Pandit08}, geological, 
as well as in meteorological and financial
time series \cite{Mandelbrot97}. 
In addition to describing their
multifractal scaling characteristics, good models should embody
their endogenous origin, i.e., the multifractal properties 
result from the explicit mechanism that 
future values of the process are (nonlinearly) influenced by the whole past history.
The endogenous self-excited nature of the generating
process indeed plays a key role in the self-organization of the complex hierarchy of the
energy cascades in turbulence, in the spatio-temporal patterns of seismicity and
in investors' decision making process leading to reflexivity in financial markets, among many others.

Multifractal processes can be considered as the next level of
generalization to the Fractional Brownian motion, which obeys
mono-scaling characterized by a single Hurst exponent $0 < H <1$,
which is itself the unique generalization of the Wiener process
$W(t)$ (with $\mathrm{E}\big[ dW \big] =0$ and $\mathrm{E}\big[ dW^2 \big]
=dt$), corresponding to the continuous time random walk with the
fixed scaling exponent $H=1/2$. 

There is a strong interest in
developing models and processes endowed with multifractal properties, 
which started from the initial models proposed by Richardson \cite{Richardson_Weather1922} and Kolmogorov \cite{Kolmogorov1941,Kolmogorov1962}. A first period can be identified, extending from 1985 to 
about 1997,  characterized by so-called cascade rules for increments of the process under analysis, which led to lognormal multifractal models \cite{jmp}.  This was followed by many extensions (see for instance  \cite{Kalisky2005,Kantelhardt2006}).
For financial time series, the direct empirical evidence of a
causal hierarchical cascade \cite{arneodicausalcasfin} motivated future developments. The
discrete hierarchical cascade approach had a number of drawbacks, such as absence 
of time dependence and discreteness of the hierarchy of scales. 
These drawbacks were solved in part by the introduction of subordinate Wiener processes expressed as functions of a non-decreasing fractal time \cite{Mandelbrot97} and of continuous multifractal cascade models \cite{Muzy2002}. 
The Multifractal Markov Switching
Model (MSM) is a significant improvement to cascade models that allows for flexible
calibration of the parameters \cite{Calvet}, but has unclear
economic or physical underpinning, and a rather artificial discrete
hierarchical structure. 

The Multifractal Random Walk (MRW)
\cite{Muzy2001, Muzy2002} is the only continuous stochastic
stationary causal process with exact multifractal properties and
Gaussian infinitesimal increments. For this, it is delicately tuned
to a critical point associated with logarithmic decay of the
correlation function of the log-increment up to an integral scale.
As a consequence, the moments of the increments of the MRW process
become infinite above some finite order, which depend on the
intermittency parameter of the model. Generalizations in terms of
log-infinitely divisible multifractal processes built as stochastic
integral of infinitely divisible 2D noise \cite{BarralMandel02}
provide more general non-linear multifractal spectra with
non-Gaussian increments \cite{BM03}. Rather than insisting on
asymptotically exact multifractal properties, the continuous-time
Quasi-Multifractal model \cite{SaichevSornette,SaichevFilimonov2007,
SaichevFilimonov2008} is based on the simple observation that
exponentials of linear long-memory processes exhibit robust
effective multifractality for all practical purpose and for a broad
range of parameters, removing the rather artificial tuning to
criticality needed in the previous models.

All these models use external innovations without explicit dependence of future values on the past history of the process.
This crucial trait makes them fundamentally unsuitable
to model the mechanisms underlying the empirical systems
mentioned above, whose multifractal fluctuations are believed to be
generated endogenously. Indeed,
turbulence is such that velocity fluctuations cascade to other
velocity fluctuations; Seismicity is predominantly powered by earthquakes
that trigger other earthquake; Financial return
fluctuations, which are weakly coupled to external news, seem mostly
driven by reflexivity \cite{Soros}. This motivates us to introduce
the \emph{Self-Excited Multifractal} (SEMF) model, as the simplest
multifractal process with self-generating properties.

\begin{figure}
  \includegraphics[width=\linewidth]{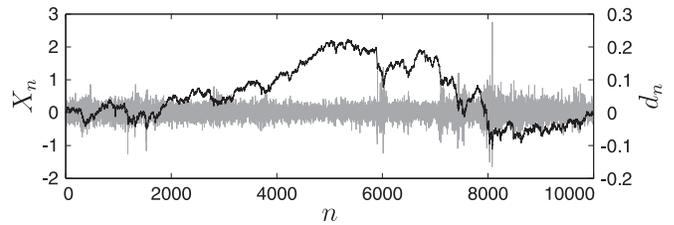}
  \caption{Time-series of increments $d_n$ (gray)
  and values $X_n=\sum_{i=0}^n d_i$ (black)
  of the discrete SEMF process defined by (\ref{eq_dn}), for $\sigma=0.01$ and
  with the power-law
  kernel~(\ref{eq_h_pow}) with  $\varphi=0.01,~h_0=0.05$.}
  \label{img_series}
\end{figure}

In discrete time $t=0, 1, 2, ..., n, ...$ (denoting without loss of
generality the time increment $\delta t \to 1$), the SEMF process reads
$X_n=\sum_{i=0}^n d_i$, where its increments are given by the
following recurrence relation
\begin{equation}\label{eq_dn}
    d_n = \sigma \xi_n\exp\left\{
    -\frac{\omega_n}{\sigma}\right\},
    \quad  \omega_n=\sum_{i=0}^{n-1}d_ih_{n-i-1} ~.
\end{equation}
The random variables $\xi_n$ represent an external noise, here
taken  i.i.d.  Gaussian with zero-mean and unit variance.
The parameter $\sigma$ sets the impact amplitude of the external
noise, as well as the dimension and scale of
 $d_n$ and $X_n$. Its value determines the time scale.  The sum in
the exponential expresses the fact that the amplitude of the next
increment of the SEMF process is strongly determined by its past
realizations, weighted by the memory kernel $h_i\geq 0, i=0, ...,
n-1$, making the dynamics ``self-excited''.
Fig.~\ref{img_series} shows a typical realization of the increments
$d_n$ defined by (\ref{eq_dn}) and of the process $X_n=\sum_{i=0}^n d_i$
with the power-law kernel
\begin{equation}\label{eq_h_pow}
  h_n=h_0 n^{-\varphi-1/2}.
\end{equation}
In the paper we will also consider the exponentially decaying kernel 
\begin{equation}\label{eq_h_exp}
  h_n=h_0\exp\left(-\phi n\right)
\end{equation}
and the constant kernel $h_n \equiv h_0$, where $\varphi$, $\phi$ and $h_0$
are positive constants.

The continuous-time version of the SEMF process is formally defined
by the stochastic integro-differential equation in the Ito
sense
\begin{equation}\label{eq_dX}
  dX(t)=\sigma\exp\left\{
  -\frac1\sigma\int\limits_{-\infty}^t h(t-t')dX(t')
  \right\}dW(t),
\end{equation}
where $dW(t)$ is the increment of the regular Wiener process and
$h(t)$ is a memory kernel function. The issue of existence of a
solution to (\ref{eq_dX}) in the strong or weak sense is considered
elsewhere. For our present purpose, it is sufficient to justify
expression (\ref{eq_dX}) as the formal notation of the discrete
formulation (\ref{eq_dn}) in the limit where the time increment
$\delta t$ goes to zero. This limit corresponds to taking $\sigma
\to 0$ appropriately ($\sigma \simeq \sqrt{\delta t}$ for the
special limiting case $h(t)=0$ retrieving the standard random walk).

The discrete SEMF process can be considered as the simplest
multifractal generalization of the GARCH process \cite{ARCH}. It can
also be viewed as a simplified time-only continuous or discrete
variant of the multifractal stress activation model introduced
earlier to model earthquake rupture processes
\cite{SornetteOuillon}. Interpreting the increments (\ref{eq_dn}) of
the SEMF process as returns of a financial time series, the random
variables $\xi_n$'s represent an external flow of news, which can
be either positive or negative, and which controls the signs of the
returns. The amplitude (or volatility) of the returns are then
determined by all past returns with a decaying weight as a function
of their distance to the present.

The previously mentioned multifractal models require stringent
conditions on their memory structure in order to exist, that is, for
their construction to converge. For instance, the MRW can be
expressed in the form (\ref{eq_dX}) but with the $dX(t')$ in the
integral in the exponential replaced by an external random Wiener
increment and, crucially, with $h(t)\sim 1/\sqrt{t}$ up to an
integral time scale and $h(t)=0$ beyond. Slower decaying kernels
make the construction non-convergent. Faster decaying kernels
destroy multifractality. The Quasi-Multifractal model
\cite{SaichevSornette, SaichevFilimonov2007} generalizes the MRW by
using a power law kernel $h(t)= h_0 /
\left(1+t/\tau\right)^{\varphi+1/2}$, and exists only for $\varphi
>0$.

In contrast, the SEMF process is much more robust and enjoy a large
domain of existence, for almost arbitrary specifications of the
memory kernel. 
Consider for instance the extreme case of a permanent
memory $h(t)\equiv h_0>0$.Then, expression (\ref{eq_dX})
becomes
\begin{equation}\label{eq_dX_const}
  dX=\sigma e^{-(h_0/\sigma)X}dW.
\end{equation}
Using the Lamperti transformation~\cite{Jeanblancetal} $Z={1 \over h_0} e^{h_0 X/\sigma}$, and Ito's lemma, equation (\ref{eq_dX_const}) leads to $dZ = {1 \over 2} {1 \over Z} dt + dW$,
whose solution is the two-dimensional Bessel process 
\begin{equation}
  B_2(t)=\sqrt{[W_1(t)]^2+[W_2(t)]^2},
\end{equation}
where $W_1(t)$ and $W_2(t)$ are two independent Wiener processes. Thus, 
\begin{equation}\label{X_const}
   X(t) = {\sigma \over h_0} \ln \left[h_0 B_2(t)\right].
\end{equation}
Since, $B_2(t)$ does not reach $0$ almost surely in
finite time and does not exhibit a finite-time singularity~\cite{Jeanblancetal}, the
process $X(t)$ is also well-behaved. 
The statistic properties of the increments of the SEMF process (\ref{X_const}) with constant kernel 
can then be obtained from the exact solution (\ref{X_const}) and will be reported elsewhere.

In general, the SEMF processes do not have stationary increments both in discrete or continuous time,
except for the trivial case $h_0=0$ which recovers the simple Wiener process. The case of a constant
memory kernel leading to the solution (\ref{X_const}) is a good illustration.
In addition to controlling the multifractality and other important properties,
the amplitude $h_0$ in (\ref{eq_h_pow}) and (\ref{eq_h_exp}) can be considered
as a ``measure of non-stationarity'' of the increments. 
Such non-stationarity might be seen as a deficiency. However, we argue that models that enforce stationarity of increments
for the convenience of their analysis may actually miss the genuine non-stationarity nature of 
many natural and social dynamics.  Let us mention in physics the case of
freely decaying two- and three-dimensional turbulence \cite{Eyink-freedecay}. In finance, there is strong evidence
supporting the view that stock market returns are non-stationary, subjected to regime shifts \cite{regimebullbear}
which can be transitory or permanent \cite{MacKenzie2008}. 

\begin{figure}
  \centering
  \includegraphics[width=0.7\linewidth]{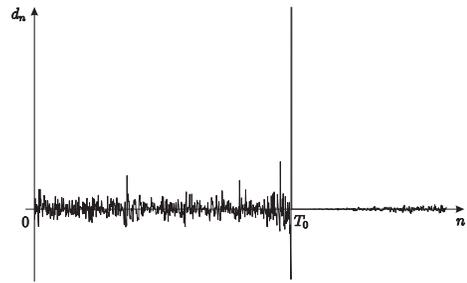}
  \caption{An extreme event in the increments of the discrete SEMF process.}
  \label{img_burst}
\end{figure}

One of the mechanisms at the source of non-stationarity in SEMF processes is the rare 
occurrence of extreme events, as illustrated by a single realization
shown in Fig.~\ref{img_burst}.  Notice the sharp growth of the volatility followed by a collapse
and a very long recovery. In the discrete version (\ref{eq_dn}), this results from rare long runs of negative innovations 
$\xi_n$ following by a sign reversal. 
The exact relation between the mean time for recovery to the size of the burst depends on 
the shape of the kernel. Simulations show that the mean time for recovery increases
faster than exponentially with the burst size for power-law kernels (\ref{eq_h_pow}). 
These extreme events are responsible for the extremely long tail of the probability density function
of increments shown in figure \ref{img_pdf} below, exemplifying the  ``dragon-king'' 
phenomenon \cite{dragonking}. Given the extreme impact of these events, it is natural
both from a mathematical point of view and motivated by empirical evidence to enrich the SEMF model
with a ``stop and resume'' rule leading to a standard renewal process
(see \cite{Feller_Probability1968}), making the renewal SEMF model having stationary increments.
In the sequel, we do not explore this extension and stick to the SEMF process  (\ref{eq_dn})
with the ``measure of non-stationarity'' $h_0<0.2$, such that 
the probability of occurrence of extreme events in time series
of duration $N=10^5$ is smaller than $10\%$. The reported statistical properties are
averaged over the realizations which do not exhibit these extreme bursts.

\begin{figure}
  \includegraphics[width=0.95\linewidth]{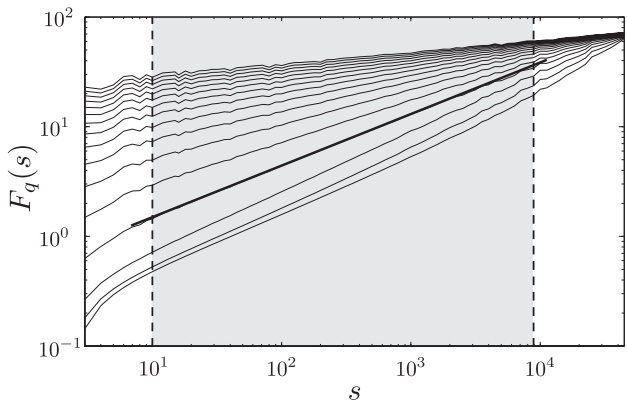}
  \caption{Log-log plot of the averaged fluctuations function given by (\ref{eq_Fq}) as a function of scale $s$,
  for $q=0.5n$, $n=1$ to $16$ (bottom to top),
  for $\sigma=1$ and power-law kernel (\ref{eq_h_pow}) with
  $\varphi=0.1$ and $h_0=0.16$. The grey rectangle delineates
  the range of scales where the fluctuation function was approximated with 
  a strict power-law $F_q(s)=K_qs^{h(q)}$. The bold line shows the power-law
  approximation for the case $q=2$.}
  \label{img_dfa_F}
\end{figure}

\begin{figure}
  \includegraphics[width=0.95\linewidth]{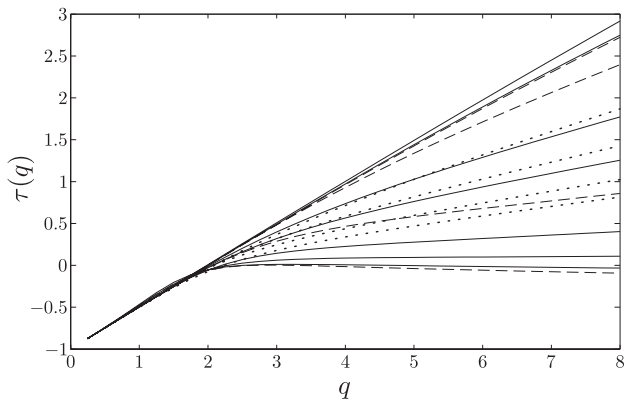}
  \caption{Multifractal scaling exponents $\tau(q)$ of the process $X_n$ for
  $\sigma=1$ and (i) power-law kernel~(\ref{eq_h_pow}) with
  $\varphi=0.01$ and $h_0=0.01,~0.06,~0.08,~0.10,~0.12,~0.14,~0.16$
  (solid lines top to bottom), (ii) exponential
  kernel~(\ref{eq_h_exp}) with  $\phi=0.01$ and
  $h_0=0.06,~0.08,~0.10,~0.12$ (dotted lines top to bottom),
  (iii) constant kernel $h\equiv h_0=0.02,~0.04,~0.06,~0.08$
  (dashed lines top to bottom).}
  \label{img_dfa_spectrum}
\end{figure}

Let consider the multifractal properties of the SEMF process.
To remove possible biases stemming from some non-stationary features
of the increments of SEMF processes, we used the multifractal detrended fluctuation 
analysis (MF-DFA)~\cite{Halvin2002} to calculate the q-th order fluctuation function
\begin{equation}\label{eq_Fq}
  F_q(s)=\left\{
    \frac1{2N_s}\sum_{\nu=1}^{2N_s}
    \big[F^2(\nu, s)\big]^{q/2}
  \right\}^{1/q},
\end{equation}
where $N_s$ is a number of segments of length $s$ within the whole time-series $X_n$ of length $N=10^5$ and $F^2(\nu,s)$ is the averaged squared residuals of the linear fit of the time series $X_n$ within the time segment $\nu$. 
To remove transient effects, we remove the first half of
the generated time series, i.e., we considered $X_n$ for $N/2\leq n\leq N$. 
Fig.~\ref{img_dfa_F} presents the q-th order fluctuation functions 
for 16 different values of the order $q$, averaged over $M=1000$ realizations of the SEMF process.
For $10 < s <N/10$, one can observe an excellent scaling regime
$F_q(s)\sim s^{h(q)}$, where the exponents $h(q)$ are the slopes in the grey area of Fig.~\ref{img_dfa_F}.
The scaling exponent $\tau(q)$ of the standard multifractal structure function
is obtained from the generalized Hurst exponent $h(q)$ by using the relationship $\tau(q)=qh(q)-1$.
The non-linear dependence of the exponents $\tau(q)$ as a function of the order $q$ shown in Fig.~\ref{img_dfa_spectrum} characterizes the multifractal properties of the process for
 different memory kernels, which include the power law (\ref{eq_h_pow}), exponential (\ref{eq_h_exp}) and the constant kernel $h_n \equiv h_0$. 
SEMF process is found to exhibit multifractal scaling over large intermediate range of scales.
The parameter $h_0$ controls the level of multifractality: increasing $h_0$
changes the spectrum from nearly monofractal for small $h_0$ values to strongly multifractal.




The probability density function (pdf) $f(d)$ of the increments
exhibits a heavy tail, as shown in Fig.~\ref{img_pdf}. Using a large
statistics of $M=10^8$ time-series of $10^3$ discrete time steps for
each of the three types of kernels (power-law, exponential and
constant), we observe three regimes for the pdf:  (i) a plateau for
small $d$'s, an intermediate asymptotic for  $2\sigma\lesssim
d\lesssim 20\sigma$ in the form of an approximate power law
\begin{equation}\label{eq_pdf_intermittent}
  f_d(d)\sim \frac1{d^{\gamma}},
\end{equation}
where $2\lesssim\gamma\lesssim6$, and
(iii) a very long tail
\begin{equation}\label{eq_pdf_big}
  f_d(d)\sim\frac1{d^\alpha \log^\beta d},
\end{equation}
where $\alpha\sim1$ and $\beta\sim2$. The specific values of the
exponents $\alpha,~\beta,~\gamma$ depend on the functional shape of
the kernel $h_n$. For instance, for the power-law
kernel~(\ref{eq_h_pow}) with $\varphi=0.01,~h_0=0.14$, we obtain
$\alpha=0.8,~\beta=4.2$ and $\gamma=4.6$. Within the financial
interpretation, choosing $\sigma \approx 1\%$ (corresponding roughly
to a time scale $\delta t \simeq 1$ day), the intermediate power law
asymptotics recovers the empirical power law distribution with an
exponent often reported in the range $3 \leq \gamma \leq 5$ for
returns in the return amplitude range $1-20\%$ \cite{Cont}. The
extreme tail (\ref{eq_pdf_big}) may correspond to the
``dragon-king'' regime \cite{dragonking}, resulting from amplifying
mechanisms leading to rare transient huge bursts. This tail can be
derived analytically using large deviation methods, as will be
reported elsewhere. Intuitively, exponentially rare streams of
negative innovations $\xi_k$ give rise to explosive growth of $d_n$,
until a positive $\xi_j$ occurs leading to a collapse of the next
increment $d_{j+1}$. The dynamics of the SEMF process is thus
characterized by quasi-stationary very long regimes punctuated by
rare bursts and collapses that play special roles, not unlike
coherent structures in turbulence or bubbles and crashes in
financial markets. 
The study of these special properties will be reported elsewhere.

\begin{figure}
  \includegraphics[width=0.95\linewidth]{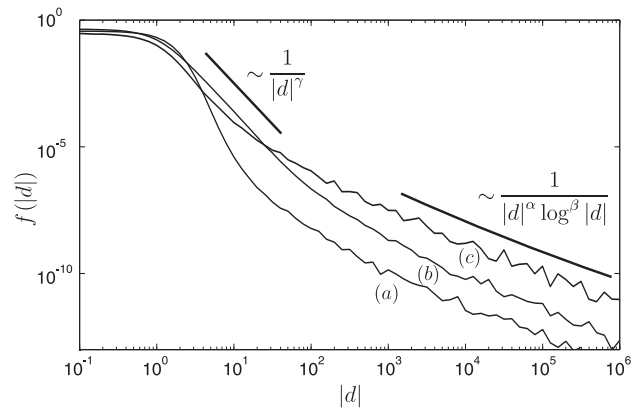}
  \caption{Probability density function of the absolute values of the
  increments $d_n$ for $\sigma=1$:
  (a) power-law kernel (\ref{eq_h_pow}) with
  $\varphi=0.1,~h_0=0.1$, (b) constant kernel
  $h_n=h_0=0.02$ and (c) exponential kernel (\ref{eq_h_exp}) with
  $\phi=0.01,~h_0=0.1$.
  \label{img_pdf}}
\end{figure}

\begin{figure}
  \includegraphics[width=0.95\linewidth]{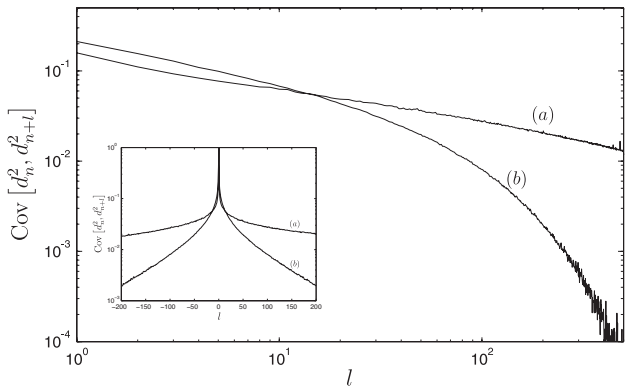}
  \caption{Covariance coefficient $\text{Cov}\left[d^2_n,d^2_{n+l}\right]$ of the squared increments 
  with $n \geq N/2$, $\sigma=1$, for
  (a) the power-law kernel (\ref{eq_h_pow}) with
  $\varphi=0.01,~h_0=0.08$ and (b) the exponential
  kernel~(\ref{eq_h_exp}) with $\phi=0.01,~h_0=0.05$.
  \revision{In this later case, for $l<150$ the covariance decays as a stretched
  exponential $\exp\left(-0.13\,x^{0.51}\right)$}. The inset shows
  the same data in \revision{log}-linear scale.
  \label{img_cov}}
\end{figure}

The SEMF process exhibits a very long dependence between the
absolute value of its increments and no dependence between the
increments themselves due to the i.i.d. random variables $\xi_n$.
The former property can be quantified by the covariance coefficient
$\text{Cov}\left[d^2_n,d^2_{n+l}\right]$. Conditioning our
estimation of this covariance coefficient by excluding the rare
occurrences of the huge explosive bursts at the origin of the very
heavy tail~(\ref{eq_pdf_big}), we obtain the results shown in
fig.~\ref{img_cov}, again after a statistical averaging over
$M=10^8$ time-series of length $10^3$.  The decay of
$\text{Cov}\left[d^2_n,d^2_{n+l}\right]$ is slower than exponential
both for slow power-law (\ref{eq_h_pow}) and
exponential~(\ref{eq_h_exp}) kernels \revision{in the range
of values $>0.001-0.01$ that is meaningful for empirical data}. In the case of the power law
kernel, $\text{Cov}\left[d^2_n,d^2_{n+l}\right] \simeq 1/l^\kappa$,
with $\kappa=0.2-0.4$, which is typical of empirical calibration of
financial returns.

As a bonus, the SEMF process exhibits the so-called \emph{leverage
effect} \cite{BouchaudLev01}, defined as a negative correlation
between past increments $d_n$ and future squared increments $d_n^2$
of the process and absence of correlation between past $d_n^2$ and
future $d_n$. The leverage effect is in general also asymmetric in
the effect of the sign of the return $d_n$: a large negative return
(loss) leads to a significant increase of volatility, while a large
positive return (gain) has smaller impact on the subsequent
volatility.  Remarkably, the SEMF process captures these effects.
Indeed, consider the return $d_n$ given by (\ref{eq_dn}) and the subsequent volatility $v_{n+1}$ that has the form: 
$$
  v_{n+1}=\exp\left\{-\frac{\omega_{n+1}}\sigma\right\}=
  \exp\left\{-\frac1\sigma\sum_{i=0}^{n-1}d_ih_{n-i} - h_0d_n\right\}.
$$
The impact of the increment $d_n$ on $v_{n+1}$ is exponential with a negative sign.
Due to the convexity of the exponential function, the relative increase ($|v_{n+1}-v_n|/v_n$)
of the volatility following a loss $d_n<0$ is larger than the relative decrease of the volatility after a gain $d_n>0$.
Moreover, the larger the previous loss/gain, the larger is the 
subsequent relative volatility increase/decrease.

\begin{figure}
  \includegraphics[width=0.95\linewidth]{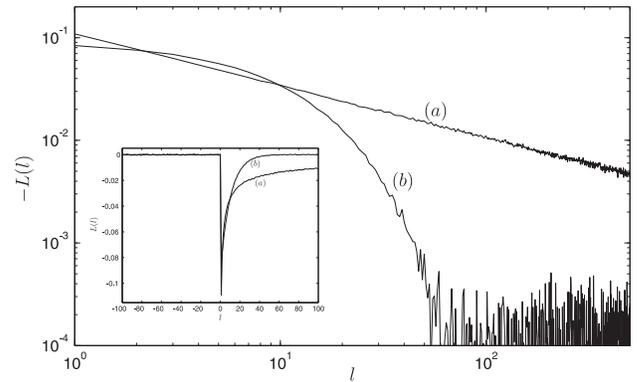}
  \caption{Leverage coefficient $L(l)$ defined in (\ref{eq_lev}) for the process $X_n$ with
  $\sigma=1$ for
  (a) the power-law kernel (\ref{eq_h_pow}) with
  $\varphi=0.01,~h_0=0.05$ and (b) the exponential
  kernel~(\ref{eq_h_exp}) with $\phi=0.1,~h_0=0.04$.
  The shape of $L(l)$ for $l>0$ is (a) power-law
  $L(l)\simeq-l^{-0.51l}$ and (b) exponential
  $L(l)\simeq-\exp(-0.1 l)$.}
  \label{img_leverage}
\end{figure}

The leverage effect can be quantified by the normalized correlation function
$L(l)=\mathrm{E}\Big[d_nd_{n+l}^2\Big] /
  \left(\mathrm{E}\Big[d_n^2\Big]\right)^{3/2}$,
which should be equal to zero for $l<0$ and negative for $l>0$
for the effect to be present. For the SEMF
process~(\ref{eq_dn}), the function $L(l)$ has the form:
\begin{equation}\label{eq_lev}
  L(l)=\frac
  {\mathrm{E}\left[\xi_n\xi^2_{n+l}
  e^{-(\omega_n+2\omega_{n+l})/\sigma}\right]}
  {\Big(\mathrm{E}\left[\xi_n^2
  e^{-2\omega_n/\sigma}\right]\Big)^{3/2}}.
\end{equation}
For $l<0$, $\omega_n$ and $\omega_{n+l}$
do not depend on the random variable $\xi_n$, leading to $L(l)=0$.
For $l>0$, we find analytically that the function~(\ref{eq_lev}) is
negative ($L(l)<0$), which proves the presence of the leverage effect
in the SEMF model. Fig.~\ref{img_leverage} illustrates this result by plotting
the function $L(l)$ obtained by averaging over a set of $M=10^8$
time-series of length $10^3$ generated by the SEMF process with the power-law
(\ref{eq_h_pow}) and exponential (\ref{eq_h_exp}) memory kernels.
One verifies the defining asymmetry of the leverage effect. It also
should be noted that the shape of function $L(l)$ for $l>0$
corresponds to the shape of memory kernel $h_n$.

Finally, we document a very interesting property of the SEMF process, namely
the breaking of statistical time reversal symmetry,
that arises naturally from its self-excitated structure, and which is related to the leverage effect.
In contrast, the MRW \cite{Muzy2001, Muzy2002} and the Quasi-Multifractal \cite{SaichevSornette,SaichevFilimonov2007, SaichevFilimonov2008} processes possess the property of statistical time reversal symmetry.
The breaking of statistical time-reversal symmetry is a stylized fact of chaotic dynamics, turbulence and 
self-organized (critical or not) systems \cite{Sornette_Critical2006}.
In particular, financial time-series are characterized by a breaking of statistical time-reversal symmetry \revision{\cite{Ramsey1996,SornetteMuzy1998}}.
In the SEMF model, the breaking of the time-reversal symmetry is already obvious from 
visual inspection of figure \ref{img_burst}, when an extreme event in the increments of the discrete SEMF process
occurs. Quantitatively, the three -point covariance function $C_3(\tau)=\mathrm{E}\big[X_n(X_{n+2\tau}-X_{n+\tau})X_{n+3\tau}\big]$ introduced by Pomeau \cite{Pomeau82} provides one of its possible diagnostic.
For $\tau=1$, we find that
\begin{equation}
 C_3(1)=\sigma^3\sum_{i=0}^n\sum_{j=0}^{n+3}
  \mathrm{E}\Big[\xi_i\xi_j\xi_{n+2}e^{-(\omega_i+\omega_j+\omega_{n+2})/\sigma}\Big]\ne0~.
\end{equation}
Because the mean under the sum is non-zero only for $j=n+2$, this leads to 
\begin{equation}\label{eq_C3}
 C_3(1)=\sigma^3\sum_{i=0}^n
  \mathrm{E}\Big[\xi_i\xi^2_{n+2}e^{-(\omega_i+2\omega_{n+2})/\sigma}\Big]~.
\end{equation}
The mean value under the sum sign has the form of the numerator 
of the expression giving the leverage coefficient (\ref{eq_lev}) with $l=n+2-i>0$, which is found negative
both in our analytical and numerical calculations. This implies that $C_3(1)<0$ 
and proves the breaking of statistical time reversal symmetry in the SEMF processes.

Summarizing, we have introduced the self-excited
multifractal (SEMF) process that exhibits strong multifractal
properties with an explicit dependence of the dynamics of the
process on both external events and internal memory. The
SEMF process enjoys all the stylized facts of self-organizing
systems such as turbulent flows, seismicity or financial markets:
multifractality, heavy tails of the distribution of increments,
absence of correlation of the signed increments and long-range
dependence in the squared increments. The ``leverage effect'' and time-reversal asymmetry are
also some of its intrinsic properties. Having the explicit feedback
of the past values on the future ones, the SEMF model is a
promising candidate for describing critical events in the self-organized
systems mentioned above.

We are grateful to Professor Alexander Saichev for fruitful
discussions.

\providecommand{\noopsort}[1]{}\providecommand{\singleletter}[1]{#1}

\end{document}